\documentclass[prb,10pt,twocolumn,showpacs,showkeys,preprintnumbers,amsmath,amssymb,superscriptaddress,aps]{revtex4-2}

\usepackage{graphicx}
\usepackage{dcolumn}
\usepackage{bm}
\usepackage{amsthm}
\usepackage{amsmath}
\usepackage{amssymb}
\usepackage{xcolor}
\usepackage{siunitx}

\DeclareSIUnit{\angstrom}{\textup{\AA}}

\begin{document}

\title{Micromagnetics of ferromagnetic/antiferromagnetic nanocomposite materials. \\ Part~II:~Mesoscopic modeling}

\author{Sergey~Erokhin}\email{s.erokhin@general-numerics-rl.de}
\affiliation{General Numerics Research Lab, Kahlaische Str.~4, D-07745 Jena, Germany}
\author{Dmitry~Berkov}
\affiliation{General Numerics Research Lab, Kahlaische Str.~4, D-07745 Jena, Germany}
\author{Andreas~Michels}
\affiliation{Department of Physics and Materials Science, University of Luxembourg, 162A~Avenue de la Faiencerie, L-1511 Luxembourg, Grand Duchy of Luxembourg}


\begin{abstract}
In the second part of this publication, we present simulation results for two three-dimensional models of Heusler-type alloys obtained by the mesoscopic micromagnetic approach. In the first model, we simulate the magnetization reversal of a single ferromagnetic (FM) inclusion within a {\it monocrystalline} antiferromagnetic (AFM) matrix, revealing the evolution of the complex magnetization distribution within this inclusion when the external field is changed. The main result of this ``monocrystalline'' model is the absence of any hysteretic behavior by the magnetization reversal of the FM inclusion. Hence, this model is unable to reproduce the basic experimental result for the corresponding nanocomposite---hysteresis in the magnetization reversal of FM inclusions with a vertical shift of the corresponding loops. To explain this latter feature, in the second model we introduce a {\it polycrystalline} AFM matrix, with exchange interactions between AFM crystallites and between the FM inclusion and these crystallites. We show that within this model we can not only reproduce the hysteretic character of the remagnetization process, but also achieve a semi-quantitative agreement with the experimentally observed hysteresis loop assuming that the concentration of FM inclusions strongly fluctuates. These findings demonstrate the reliability of our enhanced micromagnetic model and set the basis for its applications in future studies of Heusler alloys and FM/AFM nanocomposites.
\end{abstract}

\maketitle

\section{Introduction}

In the previous paper~\cite{Erokhin_PRB_2023_p1}, starting with the atomistic modeling of quasi one-dimensional (1D) systems, we have developed a novel mesoscopic micromagnetic approach for simulating materials composed of ferromagnetic (FM) inclusions in an antiferromagnetic (AFM) matrix. The need for this development is based on the discovery of strong ferromagnetism of $\rm{Ni_2MnIn}$ Heusler-type precipitates that are embedded in an AFM $\rm{NiMn}$ matrix~\cite{cakir_shell-ferromagnetism_2016,scheibel_room-temperature_2017,dincklage_annealing-time_2018}. The sizes of the FM inclusions are in the range from $5$ to $50 \, {\rm nm}$ and the magnetization curve exhibits a number of interesting features:~a vertical shift of the extracted hysteresis loop of the FM precipitates suggests a strong exchange coupling to the AFM matrix and the shape of the loop, especially its abrupt jump near zero field followed by a smooth magnetization change at much higher fields, suggests that there exist at least two different subsystems of FM inclusions.

In the following, we present a mesoscopic micromagnetic analysis of the above described Heusler system with the aim to obtain a detailed and quantitative understanding of its remagnetization processes. We remind that in the first part~\cite{Erokhin_PRB_2023_p1} we have presented atomistic and mesoscopic approaches to the micromagnetic modeling of Heusler alloys providing all necessary prerequisites for 3D mesoscopic calculations. In the present (second) part, we discuss simulation results of the full 3D models, and provide a quantitative comparison of these results to experimental data.

More specifically, in Sec.~\ref{sec2}, we simulate a single FM inclusion in a monocrystalline AFM matrix. This model, which does not reproduce the experimentally observed hysteresis, is then extended in Sec.~\ref{sec3} to include FM inclusions in a polycrystalline AFM matrix. Based on these results, we provide in Sec.~\ref{sec4} a quantitative comparison between the experimentally observed magnetization loop of Heusler-type precipitates and our simulation results, demonstrating the validity of our model.

\section{3D mesoscopic model: a single FM inclusion in a monocrystalline AFM matrix}
\label{sec2}

\begin{figure*}[ht!]
\centering                                                
\resizebox{0.90\textwidth}{!}{\includegraphics{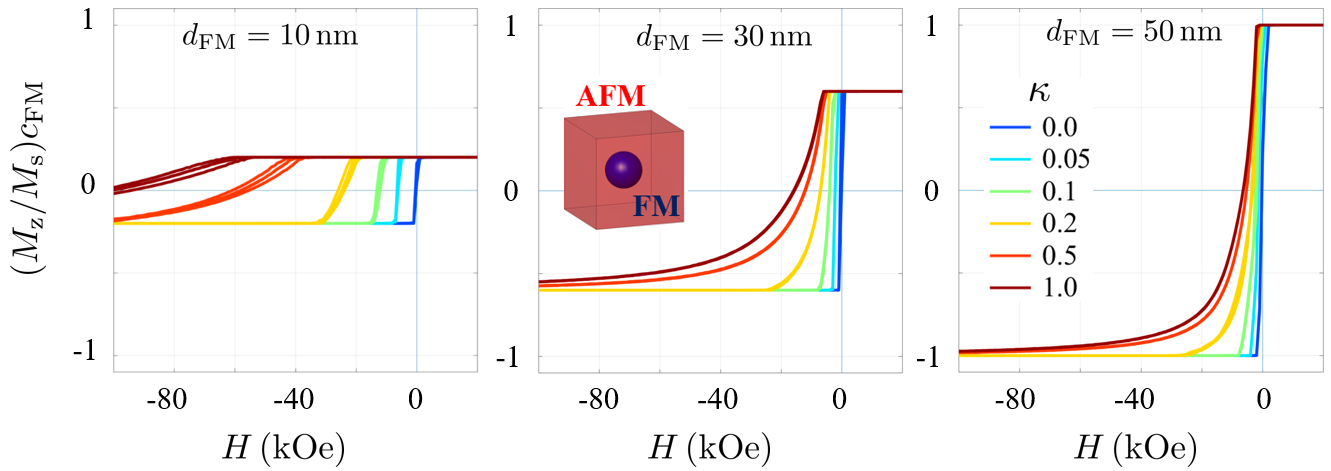}}
\caption{Simulated magnetization-reversal curves of a spherical FM inclusion (with varying size $d_{\mathrm{FM}}$) as a function of the exchange coupling $\kappa$ with the monocrystalline AFM matrix (see insets). Curves of the same color represent results for different finite-element polyhedron discretizations of the system.}
\label{figMesoSingleFMhystDfm}
\end{figure*}

The results obtained using the quasi 1D model presented in the first part~\cite{Erokhin_PRB_2023_p1} provide the framework for the next step of our study---application of our mesoscopic model to a 3D system. The need for mesoscopic simulations is based on two factors: (i)~the typical grain sizes in a nanocomposite (up to tens of nanometers) result in system sizes that are too large to be simulated using the atomistic approach, and, as we demonstrate further, (ii)~the collective nature of the magnetization-reversal process of precipitates in the polycrystalline AFM also requires to simulate systems with large sizes. Our simulations of the magnetization reversal in 3D FM/AFM structures rely on a polyhedron-based finite element micromagnetic algorithm, which we have designed specifically for the modeling of magnetic nanocomposites. The detailed description of this approach can be found in Refs.~\cite{erokhinPRB2012,michels2014jmmm}.

The primary structure of interest in this section is a spherical FM inclusion surrounded by a \textit{monocrystalline} AFM matrix (see Fig.~\ref{figMesoSingleFMhystDfm}, middle panel). In these simulations, a cubic modeling volume with a side length of $150 \, \rm{nm}$ was discretized into approximately $1.5 \times 10^5$ mesh elements, each about $3 \, \rm{nm}$ in size. This discretization allows us to study the details of the magnetization distribution of relatively small particles. Parameters of magnetic materials for both phases were presented in the first part~\cite{Erokhin_PRB_2023_p1}. We conducted simulations by varying the particle diameter $d_{\mathrm{FM}}$ of the FM inclusion between $10$ and $50 \, \rm{nm}$ and by adjusting the exchange coupling on the FM/AFM interphase boundary by varying the corresponding exchange weakening coefficient $\kappa$ between 0 (complete decoupling) and 1 (perfect exchange coupling).

The absence of magnetodipolar fields produced by the AFM phase (on a mesoscopic scale) and the negligibly small influence of the magnetodipolar field produced by the FM crystallites on the AFM matrix (in contrast to the exchange interaction) prompt us to propose a hybrid environment for micromagnetic simulations of this system. We simulate the AFM phase utilizing \textit{periodic} boundary conditions to account for the only nonlocal interaction relevant for a mesoscopic AFM---the exchange interaction. By contrast, the long-range magnetodipolar interaction inside the FM crystallite is calculated using \textit{open} boundary conditions. In this procedure the magnetodipolar interaction between the FM inclusions is neglected, which can be justified by its minor role compared to the magnetodipolar interaction of mesh elements inside the same FM inclusion. Hence, we have to compute the magnetodipolar field only within the FM inclusion, which largely reduces the computation time for the most time-consuming part of any micromagnetic simulation---the calculation of the magnetodipolar energy (the FM inclusion occupies a relatively small volume fraction of our system). This acceleration allows us to significantly extend the parameter set for our study.

Simulation results for this model presented in Fig.~\ref{figMesoSingleFMhystDfm} show magnetization-reversal curves for three FM inclusion diameters $d_{\rm FM} = 10$, $30$, and $50 \, \rm{nm}$ and various exchange weakening coefficients $\kappa$. As earlier demonstrated with the quasi~1D system, a large exchange coupling at the FM/AFM interface results in a high coercivity of the FM phase, reaching $100 \, \rm{kOe}$ for a $10$-nm-sized spherical inclusion. The exchange interaction at the FM/AFM boundary is the only mechanism connecting the phases in this one-particle model, so that when the coupling is reduced, the coercivity decreases dramatically, becoming negligible as expected for a soft FM material.

\begin{figure*}[ht!]
\centering                                                
\resizebox{0.75\textwidth}{!}{\includegraphics{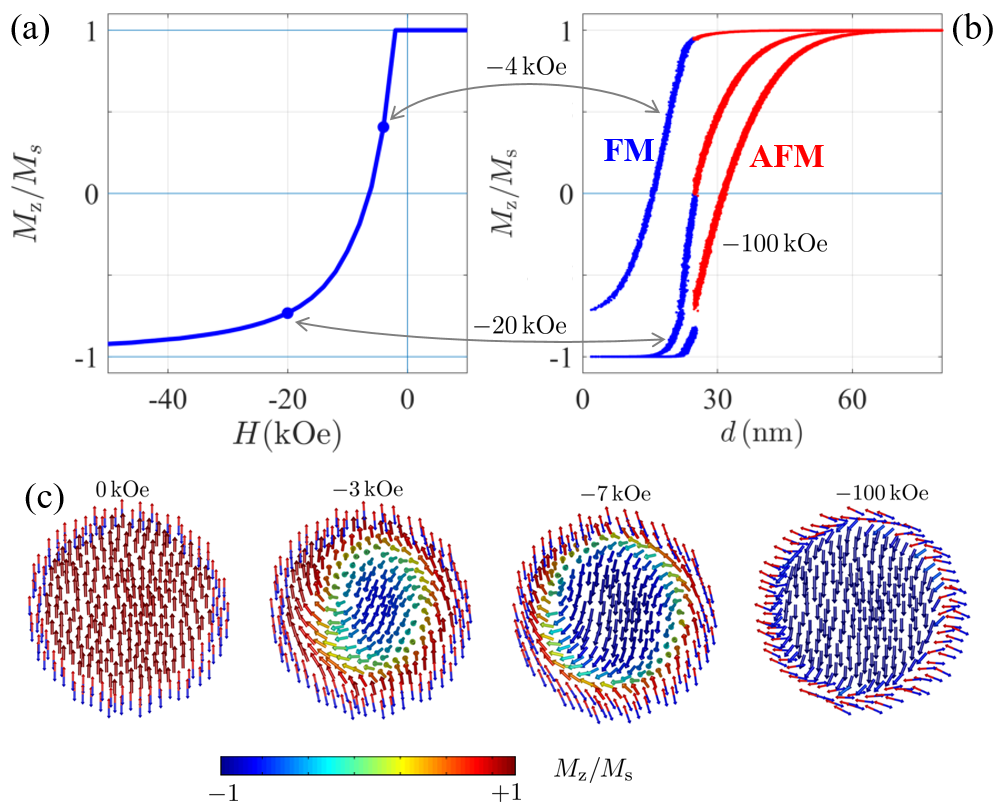}}
\caption{(a)~Magnetization-reversal curve of a 50-nm-sized spherical FM inclusion that is perfectly exchange coupled ($\kappa = 1.0$) to a monocrystalline AFM matrix. (b)~Corresponding $z$~components of magnetic moments (for one sublattice in the case of the AFM) as a function of the distance $d$ from the center of the FM inclusion. (c)~Field evolution of the magnetization distribution in the FM inclusion (embedded in the AFM matrix). The outermost red and blue shells of arrows belong to the matrix.}
\label{figMesoSingleFMDfm50nm}
\end{figure*}

Details of the remagnetization process in such a system are demonstrated in Fig.~\ref{figMesoSingleFMDfm50nm} for the example of a $50$-nm-sized particle with a perfect exchange coupling ($\kappa = 1$) with the AFM phase. Figure~\ref{figMesoSingleFMDfm50nm}(a) shows the magnetization-reversal curve of the system, and Fig.~\ref{figMesoSingleFMDfm50nm}(b) displays the $z$~component of the magnetic moments (i.e., the component parallel to the external field direction) as a function of the distance $d$ from the center of the FM particle in several external fields. For the AFM phase, only the spin-direction components of one sublattice are displayed. A notable feature of this process is a relatively swift magnetization rotation of the central part of the FM particle, attributable to the weak anisotropy of the FM material. At $H_z = -20 \, \rm{kOe}$, the majority of magnetic moments are already reversed, while the remaining ones, particularly those located in the vicinity of the interface region, form a 3D ``shell'' around the reversed inclusion kernel. The field evolution of the magnetization distribution within the FM inclusion is represented in more detail in Fig.~\ref{figMesoSingleFMDfm50nm}(c), which shows a vertical cut through the particle. Here, the formation of a shell region (where the magnetic moments are strongly coupled to the AFM matrix) and the rotation of the central part can be clearly observed.

Strong exchange coupling at the interface also results in the deviation of the AFM moments from their initial direction, as depicted in Fig.~\ref{figMesoSingleFMDfm50nm}(b). Therefore, the so-called AFM domain wall is formed in the matrix [Fig.~\ref{figMesoSingleFMDfm50nm}(c) illustrates only the AFM layer closest to the FM inclusion]. We note that the experimental observation of AFM domain structures presents a significant experimental difficulty due to the absence of magnetodipolar fields of this material at the mesoscopic scale. Thus, for AFM materials containing also FM particles, there emerges an opportunity to draw conclusions about the state of the magnetic moments of the AFM based on the magnetization distribution in the FM inclusions. Such spatial variations in the orientation of the AFM can be experimentally studied by means of the small-angle neutron scattering technique~\cite{benacchio_evidence_2019,michelsbook,bersweilerjac2022}.

The key feature of these simulation results is that \textit{hysteretic} behavior does not occur in this model, irrespective of the size of the FM inclusion or the degree of exchange weakening at the FM/AFM boundary. As it can be seen both in Fig.~\ref{figMesoSingleFMhystDfm} and \ref{figMesoSingleFMDfm50nm}(a), the magnetization of the FM inclusion follows exactly the same path for an external field ($H_z$) sweeping from $+ \infty$ to $- \infty$ and from $- \infty$ to $+ \infty$, indicating that the remagnetization process is fully reversible. This feature is due to the inability of a relatively small FM inclusion to irreversibly switch the monocrystalline AFM matrix, which has a relatively high magnetocrystalline anisotropy (see \cite{Erokhin_PRB_2023_p1}). We emphasize that neither our relatively simple model nor any more sophisticated core-shell model of a FM precipitate coupled to the AFM matrix could explain the hysteretic behavior and the nearly perfect symmetry of experimental hysteresis loops observed in~\cite{scheibel_room-temperature_2017}. The reason for this discrepancy is the fact (see above) that in such models the orientation of the AFM matrix remains nearly the same in negative and positive external fields, whereas the orientations of the magnetization within the FM inclusion are nearly opposite. This feature renders the energy minima of the complete system in high positive and high negative fields nonequivalent. Therefore, there is a need for the further improvement of our model in order to explain the experimentally observed hysteresis.


\section{Extended model: FM inclusions in a polycrystalline AFM matrix}
\label{sec3}

The reason for the absence of any hysteretic behavior for a FM inclusion embedded into a monocrystalline AFM matrix is the inability of such an inclusion to reverse the whole AFM matrix, even when the surface of the FM is perfectly exchange-coupled to the surrounding AFM. This inability is due to two system features: (i)~small concentration of FM inclusions (and only the FM fraction of the system responds to the external field), and (ii)~the very large anisotropy of the AFM material. The concentration of FM inclusions can be a subject of debate, especially when the distribution of these inclusions is strongly inhomogeneous. However, we have found that for any reasonable local concentration of the FM phase, these inclusions are not able to reverse the orientation of the entire {\it monocrystalline} AFM matrix with the nominal anisotropy of NiMn (discussed in the first part~\cite{Erokhin_PRB_2023_p1}). Taking into account that in experiment a clear hysteresis is observed, we should look for the physical explanation of the much smaller effective anisotropy of the AFM. 

In order to suggest a corresponding explanation, we remind that an analogous phenomenon is well known in the physics of ferromagnetism: the effective (volume-averaged) anisotropy $K_{\rm eff}$ of a {\it polycrystalline} ferromagnet is usually much smaller than the magnetic anisotropy of the same material in its {\it monocrystalline} state. The reason for this behavior is explained quantitatively by the Herzer model~\cite{Herzer1997}. The model takes into account that the anisotropy axes of the constituting grains in a polycrystal are usually randomly oriented, leading to random directions of the anisotropy field in each crystallite (grain). For exchange coupled grains---a normal case for a high-quality FM material---the exchange interactions between them lead to the self-averaging of the anisotropy field, resulting in a strong decrease of the volume-averaged anisotropy constant. The effect is obviously stronger in materials with a smaller average grain size $\langle d \rangle$. In fact, the effective anisotropy constant decreases rapidly as $K_{\rm eff} \sim \langle d \rangle ^6$~\cite{Herzer1997}.

Following this paradigm, we have assumed that the AFM matrix of the system studied in Ref.~\cite{cakir_shell-ferromagnetism_2016} is polycrystalline, and that the FM inclusions that are are embedded between different AFM grains exhibit a random orientation of their anisotropy planes and anisotropy axes within these planes.
We have implemented the corresponding model using a cubic simulation volume with a side length of $500 \, \mathrm{nm}$ divided into $\sim$$1.5 \times 10^5$ crystallites, each with a size of $\sim$$10 \, \mathrm{nm}$ (for both phases). The crystallites possess a polyhedron shape, the simulation volume contains no porosity, and periodic boundary conditions are implemented. Details of the microstructure generation, the discrete realization of the energy contributions, and the energy-minimization procedure can be found in Refs.~\cite{erokhinPRB2012,michels2014jmmm,ErokhinBerkov_JPCM2018}.

Magnetization reversal is simulated in frames of the Stoner-Wohlfarth model, i.e., under the assumption of a uniform magnetization of individual crystallites. The exchange coupling between the crystallites is governed by site-dependent exchange weakening coefficients $0 \leq \kappa \leq 1$, similar to the previous model. There are three kinds of couplings between the crystallites in such a two-phase system: (i)~FM/AFM coupling with $\kappa = \kappa_{\mathrm{FM-AFM}}$ (as in the previous model), (ii)~coupling between different AFM crystallites with $\kappa_{\mathrm{AFM}}$, and (iii)~coupling between FM inclusions with $\kappa_{\mathrm{FM-FM}}$. In all of the following simulations, we have set $\kappa_{\mathrm{FM-AFM}} = \kappa_{\mathrm{FM-FM}} = \kappa_{\mathrm{AFM}} = 1.0$, implying a strong exchange coupling between the corresponding phases. No magnetodipolar interaction between the crystallites is considered at the current stage.

According to measurements~\cite{scheibel_room-temperature_2017}, the total volume fraction of the precipitates in the sample is about $0.24 \, \%$, which initially motivated us to develop the one-particle model described in Sec.~\ref{sec2}. However, recent experimental observations of this class of Heusler alloys~\cite{josten2022zoom} have revealed strong spatial variations in the volume fraction of the FM phase. This finding prompted us to study the dependence of the magnetization reversal in a nanocomposite system on the volume fraction $c_{\rm FM}$ of the FM phase.

\begin{figure*}[htb!]
\centering                                                
\resizebox{1.0\textwidth}{!}{\includegraphics{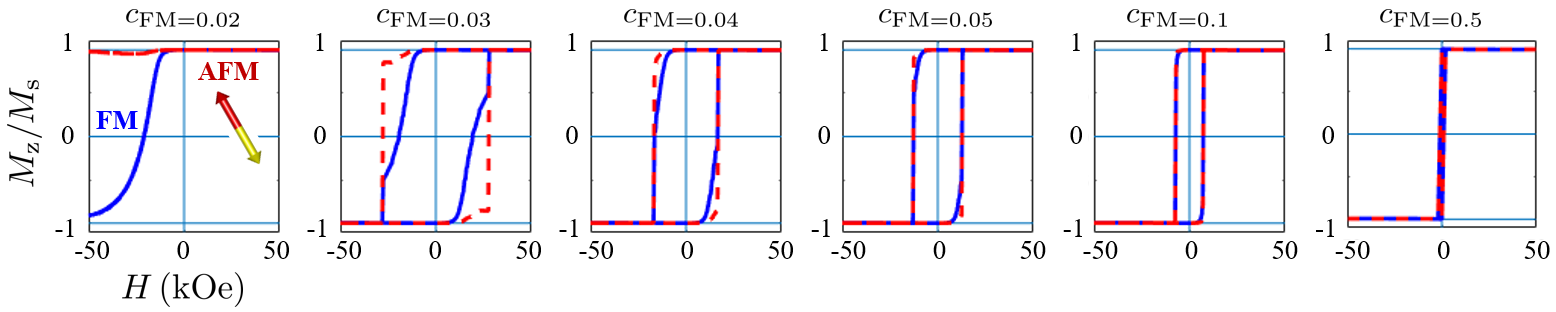}}
\caption{Simulation results for a polycrystalline FM/AFM many-particle system. Shown are magnetization-reversal curves for various volume concentrations $c_{\rm FM}$ of the FM phase (blue lines---FM phase; red dashed lines---one sublattice of the AFM phase). Both exchange couplings (FM/AFM and AFM/AFM) are perfect. The sizes of the FM inclusions and the AFM crystallites are both $10 \, \rm {nm}$.}
\label{figMesoManyFMDfm10nm}
\end{figure*}

\begin{figure*}[htb!]
\centering                                                
\resizebox{0.65\textwidth}{!}{\includegraphics{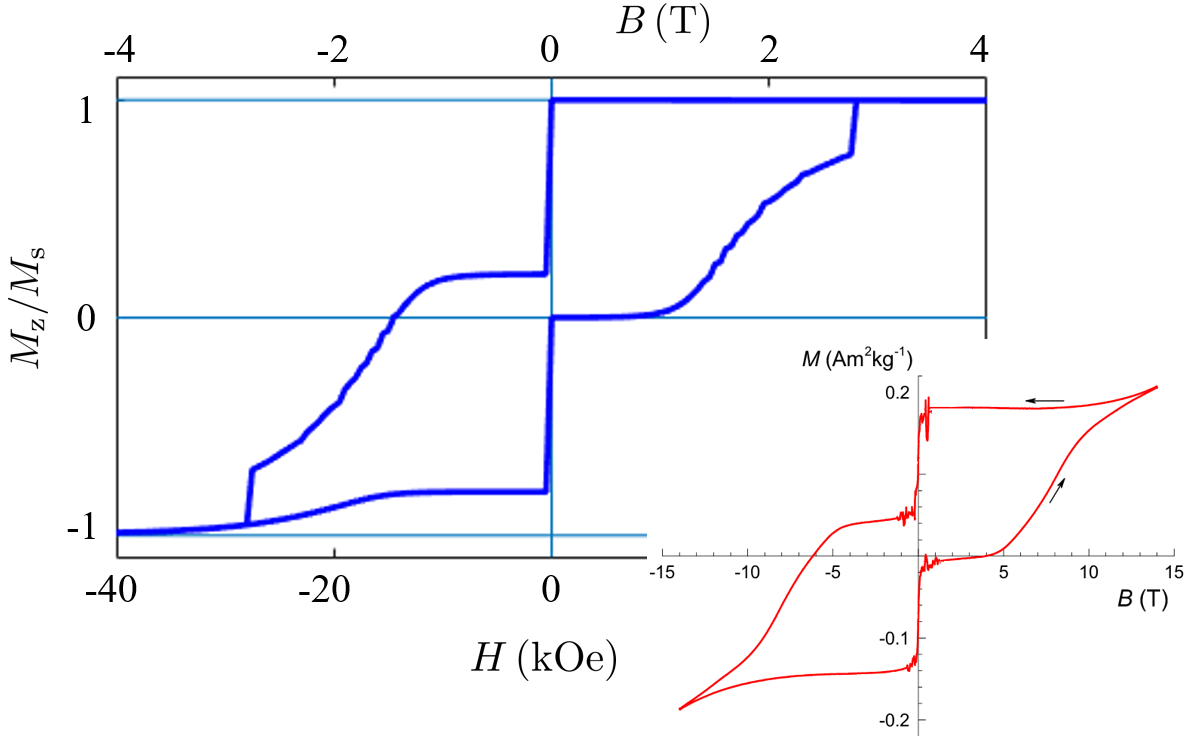}}
\caption{Composite hysteresis loop (blue line---FM phase) combined from the corresponding dependencies of systems with $c_{\rm FM} = 2, 3, 50 \, \%$ (see Fig.~\ref{figMesoManyFMDfm10nm}). The figure in the subpanel is adapted from Ref.~\cite{scheibel_room-temperature_2017} and shows the experimentally obtained magnetization loop of the FM phase in a Heusler-type alloy.}
\label{figMesoManyFMDfm10nmComposite}
\end{figure*}
  
Figure~\ref{figMesoManyFMDfm10nm} illustrates this dependence over the range of $c_{\rm FM}$ between $2$ and $50 \, \%$. As stated above, the exchange coupling between all the different crystallites is assumed to be  strong. The magnetization reversal of a nanocomposite with a low FM volume fraction ($c_{\rm FM} = 2 \, \%$) is essentially the same as for the single-particle model with the monocrystalline AFM matrix: the magnetization rotation in the system is entirely reversible (Fig.~\ref{figMesoManyFMDfm10nm}, leftmost panel) because such a low number of FM crystallites can only influence a small portion of AFM crystallites. This influence leads only to a slight deviation of the AFM magnetization curve from its maximum value. The situation changes qualitatively already at $c_{\rm FM} = 3 \, \%$, where hystereses are observed for both FM and AFM phases, indicating the complete magnetization rotation of the nanocomposite at an external field of $\sim$$- 30 \, \rm{kOe}$. Further increase of the FM volume fraction results in a decreasing coercivity, owing to the intensified interaction between the phases, until the coercivity vanishes when the volume fractions of AFM and FM are equal.

The results shown in Fig.~\ref{figMesoManyFMDfm10nm} demonstrate that the integration of spatial fluctuations of the density of FM inclusions and the polycrystallinity of the AFM matrix into our model has solved the problem of a lacking hysteretic behavior, even for small FM crystallites with a uniform magnetization. Collective interactions within the polycrystalline FM/AFM system lead to the magnetization hysteresis over a broad range of FM volume fractions. Similar simulations involving larger crystallites, which include the influence of the magnetodipolar interaction that becomes significant beyond a certain FM inclusion size [compare Fig.~\ref{figMesoSingleFMDfm50nm}(c)], are beyond the scope of this paper and will be reported in a separate publication.

\section{Model with several fractions of FM inclusions: Explanation of experimental hysteresis loops}
\label{sec4}

The results of the previous section form the basis to explain the experimentally observed hysteresis loop for a Heusler-type alloy ($\rm{Ni_2MnIn}$ precipitates embedded in an AFM $\rm{NiMn}$ matrix~\cite{scheibel_room-temperature_2017}, see the red curve in the inset of Fig.~\ref{figMesoManyFMDfm10nmComposite}). This magnetization curve clearly exhibits the characteristics of a multiphase FM subsystem, i.e., a subsystem consisting of several distinct fractions of FM inclusions: (i)~a vertical shift of the loop, (ii)~a significant drop in the magnetization at zero field, and (iii)~a broad hysteresis itself, which is typical for a system of FM particles. Here, we note that all of these features are present in the simulated magnetization-reversal curves of nanocomposites with various FM volume fractions (Fig.~\ref{figMesoManyFMDfm10nm}). Therefore, in order to show that all these features can be explained by our model, we have constructed a hysteresis curve from the already obtained results. An alternative way would be to conduct simulations by generating multiple systems with different spatial fluctuations of the FM inclusions density and then collect statistically significant characteristics of such ensembles. However, this approach would require much more computational efforts, leading to essentially the same result as described below.

An example for a hysteresis obtained from a superposition of loops shown in Fig.~\ref{figMesoManyFMDfm10nm} is displayed in Fig.~\ref{figMesoManyFMDfm10nmComposite}. Here, we demonstrate a composite loop (FM phase, blue curve), which was obtained from simulation results for systems with $c_{\rm FM} = 2, \, 3$ and $50 \, \%$ and with the assigned weights of, respectively, $0.1$\,:\,$0.5$\,:\,$0.4$. This way we have successfully modeled the experimental behavior of the FM phase under the assumption of a narrow distribution of FM grain sizes (all particles have a diameter of $\sim$10~nm) and a multimodal FM density distribution. This superposition does not only allow us to explain all qualitative aspects of the experimental magnetization curve mentioned above, but also allows us to semi-quantitatively replicate the almost horizontal plateau from $-5$~T to 0~T and the gradual magnetization reversal up to the maximum available field of $15 \, \mathrm{T}$.

\begin{figure*}[ht!]
\centering                                            
\resizebox{1.0\textwidth}{!}{\includegraphics{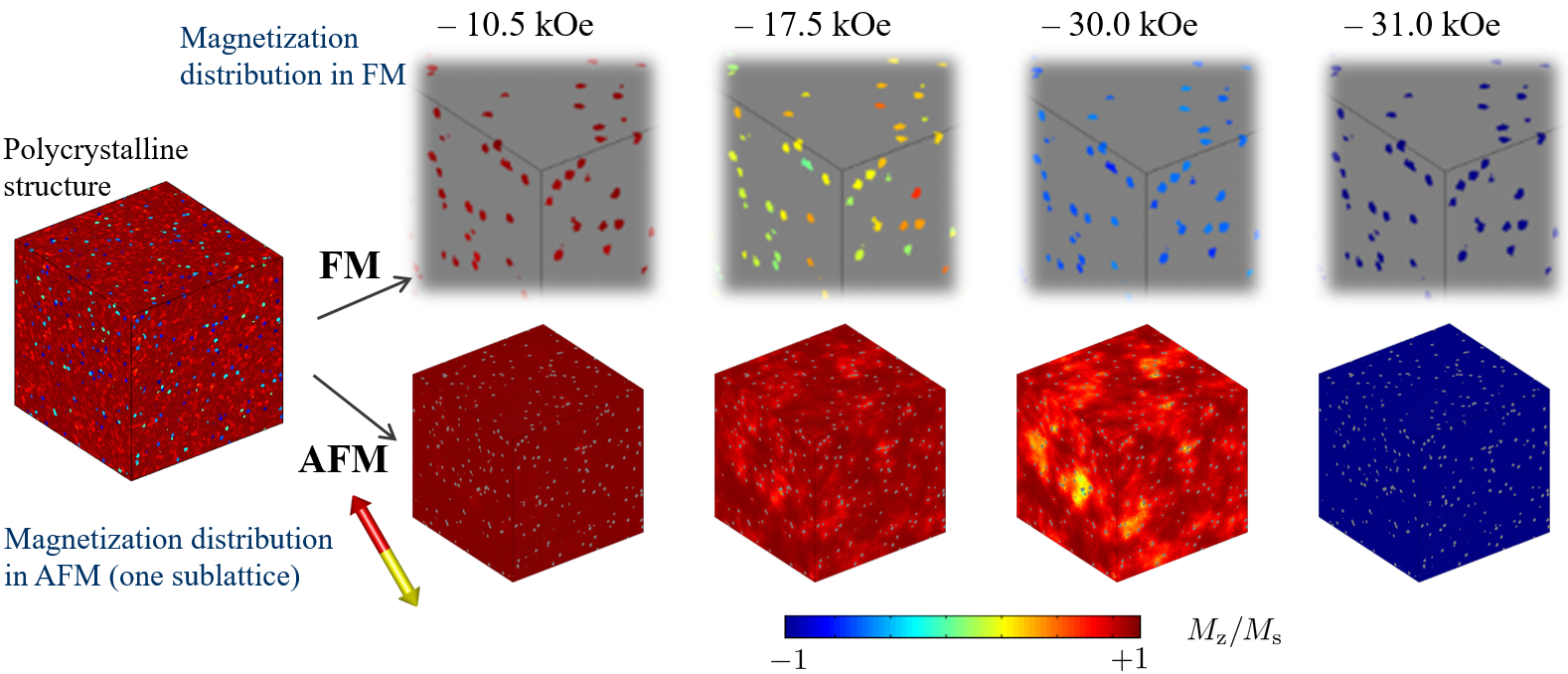}}
\caption{Evolution of the magnetization distribution in an FM/AFM polycrystalline nanocomposite with $\kappa_{\rm AFM} = 1$ and $c_{\rm FM} = 3 \, \%$. Left panel:~warm colors---AFM phases; cold colors---FM phases. Right panel:~color-coded representation of the direction of magnetic moments (upper row---FM inclusions; lower row---AFM matrix).}
\label{figMesoManyFMDfm10nmCompositeMagnDist}
\end{figure*}

The fine details of the remagnetization process in such polycrystalline nanocomposites can be revealed by analyzing the evolution of the magnetization distribution. This evolution is shown in Fig.~\ref{figMesoManyFMDfm10nmCompositeMagnDist} for a system with $\kappa_{\rm AFM} = 1$ and $c_{\rm FM} = 3 \, \%$. By relating the displayed images to the corresponding magnetization-reversal curve, we observe that both FM and AFM phases are fully magnetized at $-10.5 \,\rm{kOe}$, as it can be seen from Fig.~\ref{figMesoManyFMDfm10nm} for $c_{\rm FM} = 3 \, \%$ (in the case of the AFM phase, we mean the direction of the magnetic moments of one sublattice, with the moments of the other sublattice being antiparallel). At an intermediate field of $-17.5 \,\rm{kOe}$, a wide range of magnetization directions within the FM crystallites can be seen, while the state of the AFM matrix can be described by the onset of domain formation. A large negative field of $-30 \,\rm{kOe}$ nearly completely rotates all FM inclusions, but a clear AFM domain structure is visible. These domains are still predominantly orientated in the initial direction, i.e., opposite to the negative external magnetic field. At the final stage, the magnetic moments of all crystallites are completely reversed.

\section{Conclusion}

We have implemented a micromagnetic simulation methodology that allows to compute the magnetic response of a polycrystalline system containing ferromagnetic (FM) inclusions in an antiferromagnetic (AFM) matrix. Using this new mesoscopic simulation technique, the details of the remagnetization process in a system composed of FM crystallites embedded in an AFM matrix were revealed by simulating hysteresis curves as functions of the ferromagnetic grain size and the exchange weakening on the FM/AFM boundary. It was explicitly shown that a one-particle model employing a monocrystalline AFM matrix is incapable of explaining the experimental hysteresis results obtained on Heusler alloys. Only the inclusion of (i)~the polycrystallinity of the AFM matrix and (ii)~strong spatial fluctuations of the density of FM particles into the model explains all the qualitative features of experimental observations on these highly nontrivial systems.



\section*{Acknowledgment}

We would like to thank Nicolas Josten (University of Duisburg-Essen) for insightful discussions that greatly contributed to the research presented in this paper.

\bibliography{AFM_Heusler}

\end{document}